\documentclass[twocolumn,prb,showpacs,floatfix]{revtex4}

\usepackage{graphicx}
\usepackage{bm}
\usepackage{color}
\usepackage{amsmath}
\usepackage{amsfonts}
\usepackage{amssymb}
\usepackage{epstopdf}

\begin{document}

\preprint{APS/123-QED}

\title{Successive magnetic phase transitions in $\bm{\alpha}$-RuCl$_3$: XY-like frustrated magnet on the honeycomb lattice}

\author{Yumi Kubota$^1$}
\author{Hidekazu Tanaka$^1$}
\email{tanaka@lee.phys.titech.ac.jp}
\author{Toshio Ono$^2$}
\author{Yasuo Narumi$^3$}
\author{Koichi Kindo$^4$}

\affiliation{
$^1$Department of Physics, Tokyo Institute of Technology, Meguro-ku, Tokyo 152-8551, Japan\\
$^2$Department of Physical Science, Osaka Prefecture University, Sakai, Osaka 599-8531, Japan \\
$^3$Institute for Material Research, Tohoku University, Aoba-ku, Sendai 980-8577, Japan\\
$^4$Institute for Solid State Physics, The University of Tokyo, Kashiwa, Chiba 277-8581, Japan
}
\date{\today}

\begin{abstract}
The layered compound $\alpha$-RuCl$_3$ is composed of a honeycomb lattice of magnetic Ru$^{3+}$ ions with the $4d^5$ electronic state. We have investigated the magnetic properties of $\alpha$-RuCl$_3$ via magnetization and specific heat measurements using single crystals. It was observed that $\alpha$-RuCl$_3$ undergoes a structural phase transition at $T_{t}\,{\simeq}\,150$ K accompanied by fairly large hysteresis. This structural phase transition is expected to be similar to that observed in closely related CrCl$_3$. The magnetizations and magnetic susceptibilities are strongly anisotropic, which mainly arise from the anisotropic $g$-factors, i.e., $g_{ab}\,{\simeq}\,2.5$ and $g_c\,{\simeq}\,0.4$ for magnetic fields parallel and perpendicular to the $ab$ plane, respectively. These $g$-factors and the obtained entropy indicate that the effective spin of Ru$^{3+}$ is one-half, which results from the low-spin state. Specific heat data show that magnetic ordering occurs in four steps at zero magnetic field. The successive magnetic phase transitions should be ascribed to the competition among exchange interactions. The magnetic phase diagram for $H\,{\parallel}\,ab$ is obtained. We discuss the strongly anisotropic $g$-factors in $\alpha$-RuCl$_3$ and deduce that the exchange interaction is strongly XY-like. $\alpha$-RuCl$_3$ is magnetically described as a three-dimensionally coupled XY-like frustrated magnet on a honeycomb lattice.
\end{abstract}


\maketitle

\section{Introduction}

It is known that a honeycomb-lattice antiferromagnet with the nearest-neighbor exchange interaction undergoes a conventional magnetic ordering even for the spin-1/2 case. However, when a certain amount of second- and third-neighbor exchange interactions or a certain amount of anisotropic exchange interaction exists, the honeycomb-lattice quantum magnet exhibits an unusual ground state. In the last decade, spin-1/2 quantum magnets on honeycomb lattices have been attracting considerable attention from the viewpoints of the frustrated $J_1-J_2$ model\,\cite{Mosadeq,Ganesh,Bishop,Li2,Ganesh2,Matsuda} and the Kitaev-Heisenberg model,\cite{Singh,Chaloupka,Liu,Ye,Choi} both of which can exhibit the spin liquid state in some parameter range. $\alpha$-RuCl$_3$ appears to be a spin-1/2 honeycomb-lattice magnet.\cite{Fletcher1,Fletcher2,Kobayashi,Wang} Recently, great interest has been shown in the magnetic properties of $\alpha$-RuCl$_3$.\cite{Plumb,Sears,Majumder,Shankar} 

\begin{figure}[thb]
\begin{center}
\includegraphics[width=0.7\linewidth]{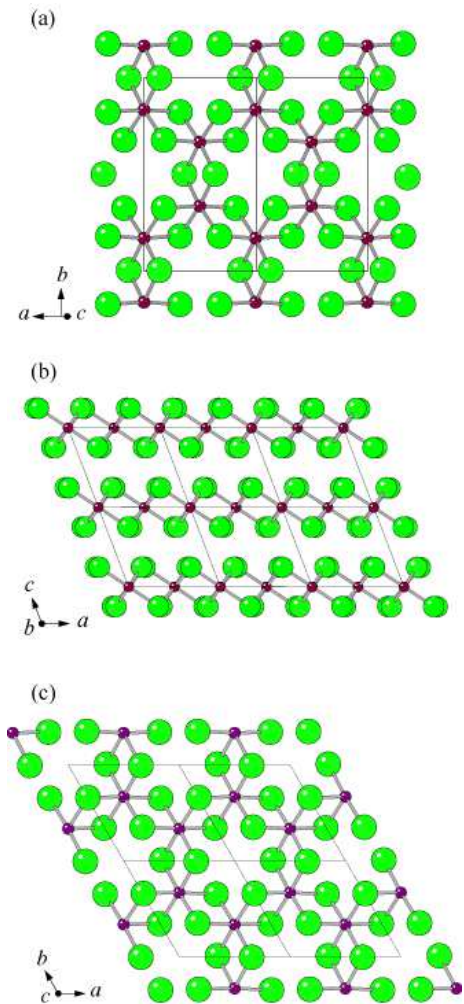}
\end{center}
\caption{Monoclinic crystal structure ($C2/m$) of $\alpha$-RuCl$_3$ at room temperature (a) viewed perpendicular to the $ab$ plane and (b) viewed along the $b$ axis. Small purple and large green spheres are Ru$^{3+}$ and Cl$^-$ ions, respectively. Thin solid lines denote the chemical unit cells. (c) Trigonal crystal structure ($R{\bar 3}$) of CrCl$_3$ in the low-temperature phase viewed along the $c$ axis, which is expected to be the same as the low-temperature crystal structure of $\alpha$-RuCl$_3$.} 
\label{structure}
\end{figure}

$\alpha$-RuCl$_3$ has a layered structure. The crystal structure was first reported to be trigonal, $P3_112$,\cite{Fletcher1,Fletcher2} but later it was found to be monoclinic, $C2/m$,\cite{Brodersen} which is the same as the room-temperature crystal structure of CrCl$_3$.\cite{Cable,Morosin} Figures~\ref{structure}(a) and (b) show the crystal structure of $\alpha$-RuCl$_3$.
The crystal structure is composed of RuCl$_6$ octahedra, which are linked in the $ab$ plane by sharing edges. Magnetic Ru$^{3+}$ ions with the $4d^5$ electronic state form a slightly distorted honeycomb lattice. It has been reported that $\alpha$-RuCl$_3$ undergoes magnetic ordering at $T_{\rm N}\,{=}\,13-15.6$ K.\cite{Fletcher2,Kobayashi,Wang} However, little is known about the magnetic properties of $\alpha$-RuCl$_3$.  

When Ru$^{3+}$ has the high-spin state, the total angular momentum and total spin are given by $L\,{=}\,0$ and $S\,{=}\,5/2$, respectively; thus, the magnetic moment is given by the spin only. Consequently, the exchange interaction and $g$-factor become isotropic. On the other hand, when Ru$^{3+}$ has the low-spin state, the orbital and spin states are described by $l\,{=}\,1$ and $S\,{=}\,1/2$, respectively. In this case, the magnetic moment is given by the effective spin-1/2, which is composed of the orbital angular momentum and true spin. In general, the exchange interaction and $g$-factor for the effective spin are fairly anisotropic.  

In this paper, we present the results of magnetization and specific heat measurements on $\alpha$-RuCl$_3$. It was observed that the magnetic susceptibilities in $\alpha$-RuCl$_3$ are strongly anisotropic, which indicates the low-spin state of Ru$^{3+}$ together with small entropy. It was found that the magnetic ordering occurs in multiple steps, which we consider to be due to competition between exchange interactions. This paper is organized as follows. Experimental procedure is described in Section II. Experimental results are given in Section III. The exchange interaction, the $g$-factor for the low-spin state of Ru$^{3+}$ and the phase diagram are discussed in Section IV. Section V is devoted to a conclusion.

\section{Experimental details}

Single crystals of RuCl$_3$ were grown from a melt by the vertical Bridgman technique. Fine-grained RuCl$_3$ was dehydrated in a quartz tube at $100\,^{\circ}$C for three days. The temperature of the center of the furnace was set at $1100\,^{\circ}$C and the quartz tube was moved downward in the furnace at a rate of 3\,mm/h over 80\,h. The crystals obtained were black and had wide surfaces parallel to the $ab$ plane. The crystals were soft and easily bent like foil.

Magnetization measurements were performed using a SQUID magnetometer (Quantum Design MPMS XL) in the temperature range $1.8\,\mathrm{K} \leq T \leq 100\,\mathrm{K}$ in magnetic fields of up to 7 T. Magnetic fields were applied parallel and perpendicular to the $ab$ plane. High-field magnetization measurement in a magnetic field of up to 57.5 T was performed at 4.2 and 1.3 K using an induction method with a multilayer pulse magnet at the Institute for Solid State Physics, University of Tokyo. The absolute value of the high-field magnetization was calibrated with the magnetization measured by the SQUID magnetometer. The specific heat was measured down to 0.36 K in magnetic fields of up to 9 T using a physical property measurement system (Quantum Design PPMS) by the relaxation method.

\section{Experimental Results}
\begin{figure}[t]
\begin{center}
\includegraphics[width=0.75\linewidth]{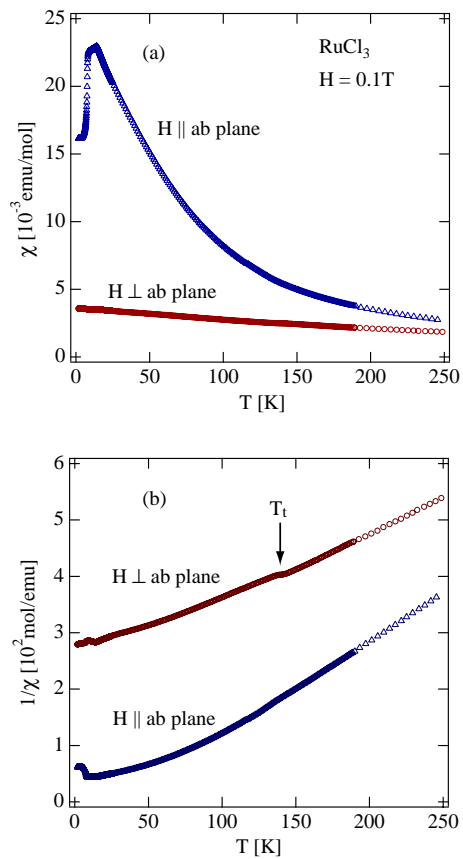}
\end{center}
\caption{
Temperature dependence of (a) magnetic susceptibilities ${\chi}\,{=}\,M/H$ and (b) inverse magnetic susceptibilities in $\alpha$-RuCl$_3$ measured at $H\,{=}\,0.1$\,T for $H\,{\parallel}\,ab$ and $H\,{\perp}\,ab$. The arrow indicates the structural phase transition temperature $T_{\rm t}$. 
} 
\label{chi_1}
\end{figure}

Figures~\ref{chi_1}(a) and (b) show the temperature dependence of the magnetic susceptibilities and inverse susceptibilities measured for a magnetic field $H$ parallel and perpendicular to the $ab$ plane. The magnetic susceptibility for $H\,{\parallel}\,ab$ plane is much larger than that for $H\,{\perp}\,ab$ plane. The strongly anisotropic susceptibility is mainly attributed to the strongly anisotropic $g$-factor, which results from the low-spin state of Ru$^{3+}$. The anisotropy of the $g$ factor is discussed in detail in the next section.

\begin{figure}[t]
\begin{center}
\includegraphics[width=0.75\linewidth]{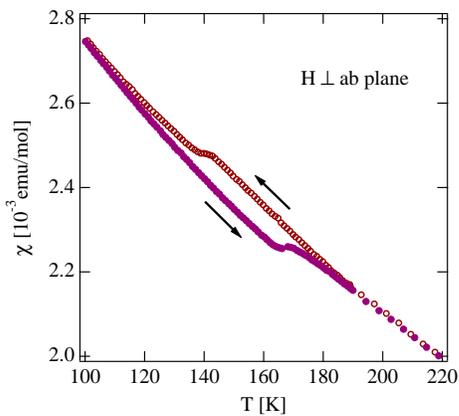}
\vspace{-2mm}
\end{center}
\caption{Hysteresis in magnetic susceptibility observed around $T\,{=}\,150$ K for $H\,{\perp}\,ab$.} 
\label{chi_2}
\end{figure}

\begin{figure}[h]
\begin{center}
\includegraphics[width=0.75\linewidth]{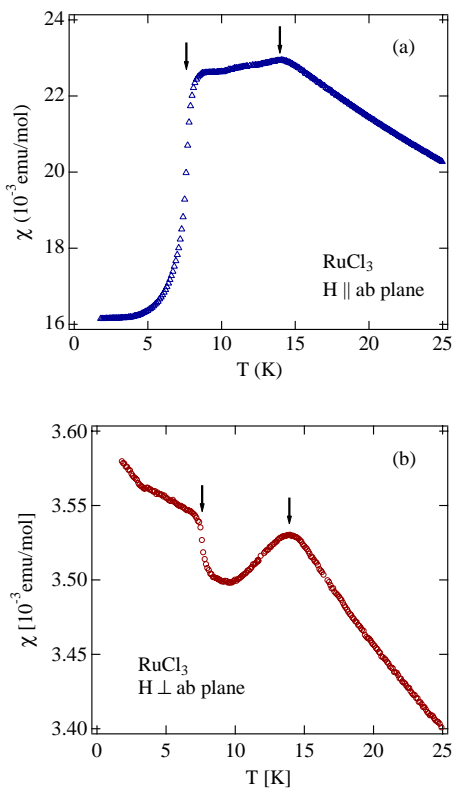}
\vspace{-2mm}
\end{center}
\caption{(a) Low-temperature magnetic susceptibilities in $\alpha$-RuCl$_3$ measured at $H\,{=}\,0.1$\,T for (a) $H\,{\parallel}\,ab$ and (b) $H\,{\perp}\,ab$. Arrows indicate the anomalies caused by magnetic phase transitions.} 
\label{chi_3}
\end{figure}

In the inverse susceptibility for $H\,{\perp}\,ab$ plane, a discontinuous change is observed at $T_{\rm t}\,{\simeq}\,150$ K. Figure~\ref{chi_2} shows an enlargement of the magnetic susceptibility for $H\,{\perp}\,ab$ plane around $T_{\rm t}$. Hysteresis is clearly observed at the transition temperatures, $T_{\rm t}\,{=}\,141$ and 167 K upon cooling and heating, respectively. This anomaly  in the susceptibility is ascribed to the structural phase transition. Although we performed X-ray crystal analysis below $T_{\rm t}$, we could not determine the low-temperature structure because sharp X-ray spots were not observed owing to the softness of the crystal. In the closely related compound CrCl$_3$, a structural phase transition from the monoclinic structure ($C2/m$) to the trigonal structure ($R{\bar 3}$) takes place at $T_{\rm t}\,{\simeq}\,240$ K.\cite{Morosin} Thus, it is likely that the low-temperature structure is the same as the low-temperature structure of CrCl$_3$, which is shown in Fig.~\ref{structure}(c). 
It is noted that the specific heat data show no sharp anomaly at the structural phase transition temperature $T_{\rm t}\,{\simeq}\,150$ K. We consider that this is because the relaxation method used in this work is less sensitive to the first order phase transition with large latent heat.

\begin{figure}[t]
\begin{center}
\includegraphics[width=0.75\linewidth]{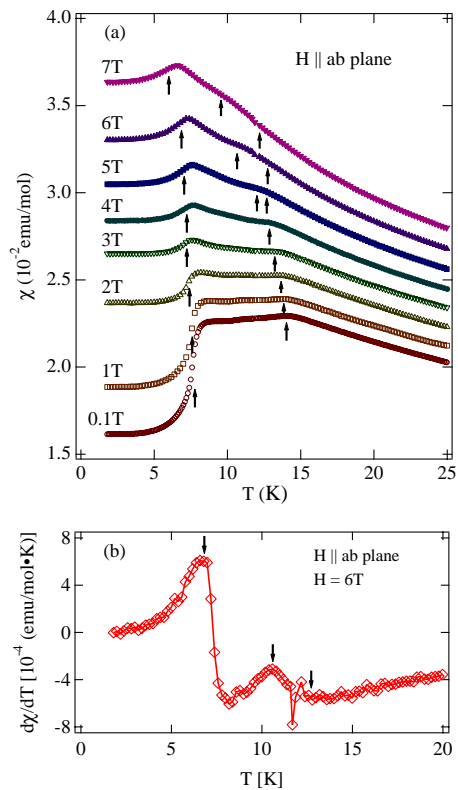}
\end{center}
\caption{Low-temperature magnetic susceptibilities in $\alpha$-RuCl$_3$ measured at various magnetic fields for $H\,{\parallel}\,ab$. The susceptibility data are shifted upward by multiples of  $2\,{\times}\,10^{-3}$ emu/mol. Arrows indicate the transition temperatures determined from the specific heat anomaly. (b) Derivative of ${\chi}$ with respect to $T$ measured at $H\,{=}\,6$ T. Arrows indicate the transition temperatures determined from the specific heat anomaly, which are close to those giving local maxima or minima of $d{\chi}/dT$.} 
\label{chi_4}
\end{figure}

Figure~\ref{chi_3} shows the low-temperature magnetic susceptibilities measured for $H\,{\parallel}\,ab$ and $H\,{\perp}\,ab$. Clear anomalies indicative of magnetic phase transitions are observed at $T\,{=}\,13.9$ and 7.6 K in both susceptibility data. These phase transitions are confirmed by performing specific heat measurements as shown below. Our low-temperature magnetic susceptibilities are consistent with those reported recently by Sears {\it et al.}\cite{Sears} and Majumder {\it et al.}\cite{Majumder} However, we observed that for $H\,{\perp}\,ab$, the magnetic susceptibilities in some $\alpha$-RuCl$_3$ samples decrease at 7.6 K with decreasing temperature in contrast to the behavior shown in Fig.~\ref{chi_3}(b).

Figure~\ref{chi_4}(a) shows the low-temperature magnetic susceptibility ${\chi}\,{=}\,M/H$ measured at various magnetic fields for $H\,{\parallel}\,ab$. Arrows indicate the transition temperatures determined from the specific heat anomaly. These transition temperatures are close to those giving local maxima or minima of $d{\chi}/dT$, as shown in Fig.~\ref{chi_4}(b). The spike anomaly at 12 K is due to an instrumental problem. With increasing magnetic field, the transition temperatures decrease, and for $H\,{>}\,4$ T, the high-temperature transition splits into two transitions.

\begin{figure}[t]
\begin{center}
\includegraphics[width=0.75\linewidth]{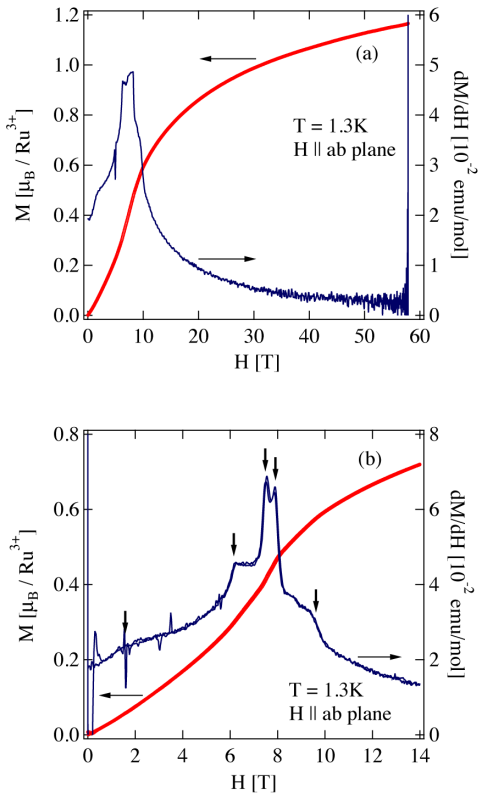}
\end{center}
\caption{Magnetic field dependence of the magnetization $M$ and its field derivative $dM/dH$ measured at 1.3 K for $H\,{\parallel}\,ab$ in $\alpha$-RuCl$_3$ for the highest fields of (a) 57.5 T and (b) 14 T. Arrows indicate the transition fields.} 
\label{MH_perp}
\end{figure}
\begin{figure}[h]
\begin{center}
\includegraphics[width=0.75\linewidth]{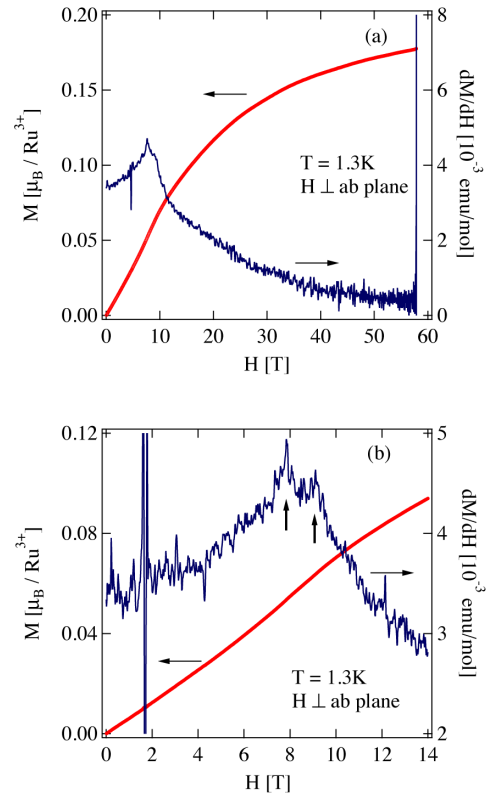}
\end{center}
\caption{Magnetic field dependence of the magnetization $M$ and its field derivative $dM/dH$ measured at 1.3 K for $H\,{\perp}\,ab$ in $\alpha$-RuCl$_3$ with the highest fields of (a) 57.5 T and (b) 20 T. Arrows indicate the transition fields.} 
\label{MH_para}
\end{figure}

Figures~\ref{MH_perp} and \,\ref{MH_para} show the magnetic field dependence of magnetization and its field derivative $dM/dH$ measured at 1.3 K for $H\,{\parallel}\,ab$ and $H\,{\perp}\,ab$, respectively. In both figures, the upper and lower panels show data taken with the highest magnetic fields of 57.5 and 14 T, respectively. The absolute values of the magnetization for $H\,{\parallel}\,ab$ and $H\,{\perp}\,ab$ are considerably different, as observed in the magnetic susceptibilities. This is ascribed to the strongly anisotropic $g$-factors. Extrapolating the magnetization curves to higher fields, we estimate the $g$-factors to be $g_{ab}\,{=}\,2.5\,{\pm}\,0.2$ and $g_c\,{=}\,0.40\,{\pm}\,0.03$. As shown by arrows in Figs.\,\ref{MH_perp} and \,\ref{MH_para}, some anomalies in $dM/dH$ indicative of field-induced phase transitions are observed at $H\,{=}\,1.6, 6.2, 7.5, 7.9$ and 9.6 T for $H\,{\parallel}\,ab$, and at $H\,{=}\,7.9$ and 9.1 T for $H\,{\perp}\,ab$. For $H\,{\parallel}\,ab$, the field-induced phase transition occurs in many steps. However, a distinct transition to saturation is not observed for either field direction, despite the sufficiently low temperature of 1.3 K. This indicates that the total spin is not conserved. Because $\alpha$-RuCl$_3$ is considered to be a localized spin system at helium temperatures, a strong antisymmetric interaction such as the Dzyaloshinskii-Moriya (DM) interaction may be responsible for the smearing of the saturation transition.

\begin{figure*}[t]
\begin{center}
\includegraphics[width=1.0\linewidth]{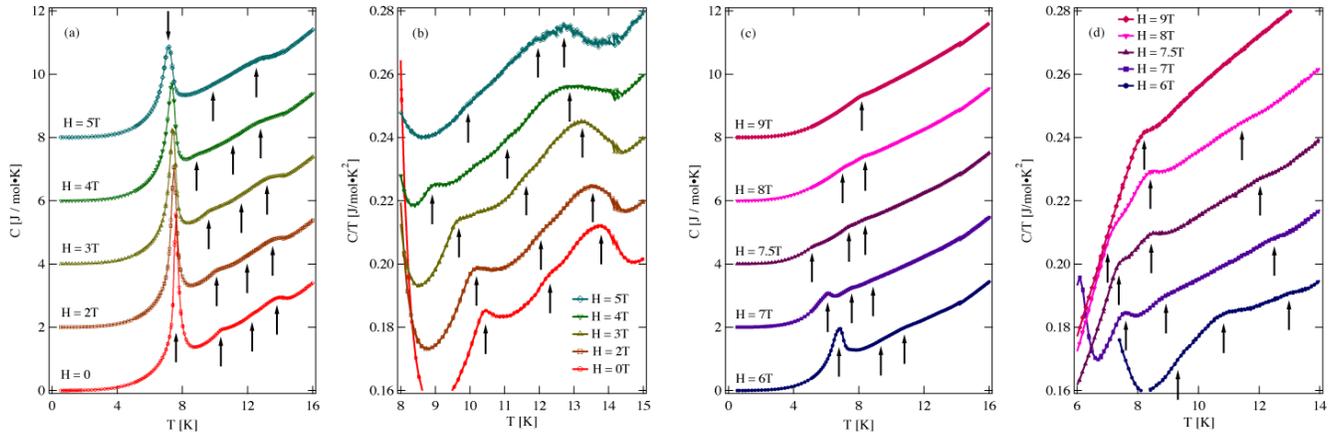}
\end{center}
\caption{Low-temperature specific heat $C$ and $C/T$ in $\alpha$-RuCl$_3$ measured at various magnetic fields for $H\,{\parallel}\,ab$. The specific heat data and $C/T$ are shifted upward by multiples of 2 J/mol${\cdot}$K and 0.2 J/mol${\cdot}$K$^2$, respectively. Arrows show the anomalies indicative of magnetic phase transitions.} 
\label{heat}
\end{figure*}

Figure~\ref{heat} shows the temperature dependence of specific heat $C$ and $C/T$ measured at various magnetic fields for $H\,{\parallel}\,ab$. The anomaly in $C/T$ near 14 K is due to the instrumental problem. Our specific heat data are consistent with those reported by Majumder {\it et al.}\cite{Majumder}. At zero magnetic field, a sharp peak is observed at $T_{\rm N4}\,{=}\,7.6$ K. Above 7.6 K, three small anomalies are observed at $T_{\rm N3}\,{=}\,10.4$ K, $T_{\rm N2}\,{=}\,12.3$ K and $T_{\rm N1}\,{=}\,13.8$ K as indicated by arrows. We consider that these anomalies arise from magnetic phase transitions, because they are sharper than those owing to the short-range spin correlation. The present result shows that the magnetic ordering occurs in four steps. The transitions at $T_{\rm N1}$ and $T_{\rm N4}$ are also clearly observed in the magnetic susceptibility as shown in Fig.~\ref{chi_3}.

\begin{figure}[h]
\begin{center}
\includegraphics[width=0.75\linewidth]{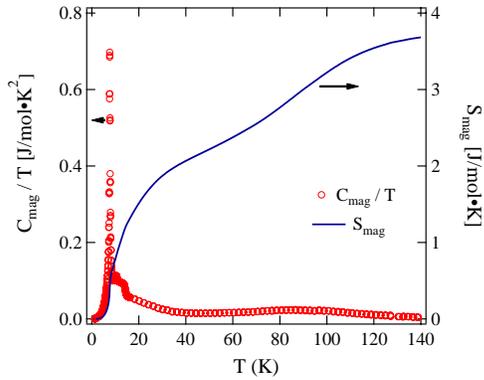}
\end{center}
\caption{Temperature dependence of $C_{\rm mag}/T$ in $\alpha$-RuCl$_3$ measured at zero magnetic field. The solid curve represents the magnetic entropy $S_{\rm mag}$.} 
\label{entropy}
\end{figure}

\begin{figure}[h]
\begin{center}
\includegraphics[width=0.75\linewidth]{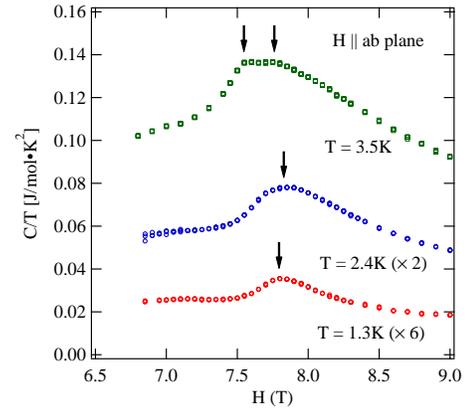}
\end{center}
\caption{Field scans of the specific heat in $\alpha$-RuCl$_3$ measured at 1.3, 2.4 and 3.5 K for $H\,{\parallel}\,ab$. Arrows show the anomalies indicative of magnetic phase transitions.} 
\label{heat_3}
\end{figure}

The magnetic specific heat $C_{\rm mag}$ was evaluated using the specific heat of nonmagnetic ScCl$_3$, which has a similar layered crystal structure to RuCl$_3$.\cite{Fjellvag} Figure~\ref{entropy} shows the temperature dependence of $C_{\rm mag}/T$ and magnetic entropy $S_{\rm mag}$ at zero magnetic field. $C_{\rm mag}/T$ has a broad maximum around 85 K, which is interpreted to be caused by the short-range spin correlation. The magnetic entropy obtained at 140 K ($\,{<}\,T_{\rm t}$) is $S_{\rm mag}\,{=}\,3.8$ J/mol\,$\cdot$\,K, which is approximately two-third of $R\ln 2\,{=}\,5.76$. This small magnetic entropy is consistent with the low-spin state of Ru$^{3+}$ with effective spin-1/2. 

As shown in Fig.~\ref{heat}, with increasing magnetic field for $H\,{\parallel}\,ab$, the peak heights of the anomalous specific heat for all the phase transitions decrease and all the transition temperatures shift gradually towards the low-temperature side. The lowest transition temperature $T_{\rm N4}$ decreases rapidly above 7 T. This can be observed in the field scans of the specific heat at three temperatures below 4 K, as shown in Fig.~\ref{heat_3}. The peak around 7.8 T, indicative of a magnetic phase transition, is interpreted to be connected to the transition line for $T_{\rm N4}$ for $H\,{\leq}\,7$ T. The high-temperature transition $T_{\rm N1}$ splits into two transitions above $H\,{=}\,5$ T. 

\begin{figure}[h]
\begin{center}
\includegraphics[width=0.75\linewidth]{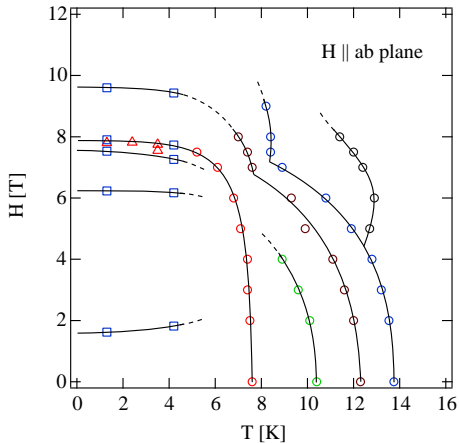}
\end{center}
\caption{Magnetic phase diagram for $H\,{\parallel}\,ab$. The circular and triangular symbols are transition points determined from the temperature and field dependences of specific heat, respectively, and the rectangular symbols are those determined from the magnetization process. Solid and dashed lines are visual guides.} 
\label{phase}
\end{figure}

The transition data obtained for $H\,{\parallel}\,ab$ are summarized in Fig.~\ref{phase}. Four ordered phases exist at zero magnetic field and six ground states exist in magnetic fields. The features of  $\alpha$-RuCl$_3$ observed in the present measurements are strongly anisotropic magnetic properties and many ordered phases. In the next section, we discuss these features.

\section{Discussion}
\subsection{Effective exchange model and $g$-factor}
In this subsection, we discuss the strongly anisotropic magnetic properties and derive an effective model that describes the low-temperature and low-energy magnetic properties of $\alpha$-RuCl$_3$ in accordance with the effective model of Co$^{2+}$ in an octahedral environment.\cite{Abragam,Lines} Because the magnetic entropy obtained from the specific heat data indicates that the effective spin of Ru$^{3+}$ is one-half, it is natural to assume that the five electrons in the $4d$ orbitals of Ru$^{3+}$ are in the low-spin state owing to the strong crystalline fields from surrounding Cl$^-$ ions. 

In the low-spin state, all five electrons in the $4d$ orbitals occupy the $d{\epsilon}$ orbital. Because the matrix elements of the orbital angular momenta $l_{d{\epsilon}}^x, l_{d{\epsilon}}^y$ and $l_{d{\epsilon}}^z$ with respect to the orbital states ${\phi}_{\xi}, {\phi}_{\eta}$ and ${\phi}_{\zeta}$ for the $d{\epsilon}$ orbital are given by changing the sign of those for the orbital angular momenta $l_{p}^x, l_{p}^y$ and $l_{p}^z$ with respect to the orbital states ${\phi}_{x}, {\phi}_{y}$ and ${\phi}_{z}$ for the $p$ orbital, respectively, we can replace ${\bm l}_{d{\epsilon}}$ by $-{\bm l}$ with $l\,{=}\,1$. The spin-orbit coupling of these electrons is expressed as
\begin{eqnarray}
{\cal H}_{\rm so}=\sum_{i\,{=}\,1}^{5} \frac{g{\mu_{\rm B}}^2Z}{r_i^3}({\bm l}_{d{\epsilon},i}\,{\cdot}\,{\bm s}_i)=-\sum_{i\,{=}\,1}^{5} \frac{g{\mu_{\rm B}}^2Z}{r_i^3}({\bm l}_{i}\,{\cdot}\,{\bm s}_i).
\label{spin-orbit}
\end{eqnarray} 
For the three electrons with up spin, their orbital angular momenta cancel out, $\sum_{i\,{=}\,1}^3{\bm l}_i\,{=}\,0$. For the other two electrons with down spin, their spin ${\bm s}_i\ (i\,{=}\,4\ {\rm and}\ 5)$ is expressed using the total spin ${\bm S}$ with $S\,{=}\,1/2$ as ${\bm s}_i\,{=}\,-{\bm S}$. For these reasons, the spin-orbit coupling of eq.\,(\ref{spin-orbit}) is written as
\begin{eqnarray}
{\cal H}_{\rm so}=g{\mu_{\rm B}}^2Z\left\langle\frac{1}{r^3}\right\rangle ({\bm l}\cdot{\bm S})={\lambda}({\bm l}\cdot{\bm S}).
\label{spin-orbit2}
\end{eqnarray}
The coupling constant $\lambda$ is positive and its magnitude has been reported to be ${\lambda}\,{\simeq}\,1000$\,cm$^{-1}$.\cite{Geschwind} When the $p$ orbital of the surrounding Cl$^-$ is mixed with the $4d$ orbitals of Ru$^{3+}$, the matrix elements of the angular momentum $\bm l$ are reduced. This effect is expressed by replacing ${\bm l}$ with $k{\bm l}$ with $0<k\leq 1$. 

\begin{figure}[b]
\begin{center}
\includegraphics[width=0.8\linewidth]{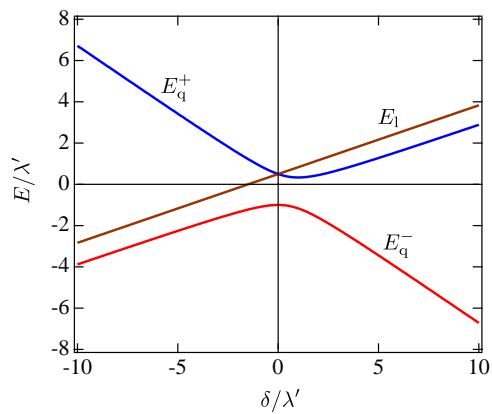}
\end{center}
\caption{Energy levels of $E_{\rm l}$ and $E_{\rm q}^{\pm}$ as a function of ${\delta}/{\lambda}^{\prime}$.} 
\label{Kramers}
\end{figure}

The orbital state of the low-spin state of Ru$^{3+}$ in an octahedral environment is triply degenerate. The orbital degeneracy can be lifted by the spin-orbit coupling and the trigonal crystalline field, which are written as
\begin{eqnarray}
{\cal H}^{\prime}={\lambda}^{\prime}({\bm l}\cdot{\bm S})+{\delta}\left\{(l^z)^2-2/3\right\},
\label{perturbation}
\end{eqnarray}
where ${\lambda}^{\prime}=k{\lambda}$, and the second term represents the energy of the trigonal crystalline field. When the RuCl$_6$ octahedron is trigonally compressed, ${\delta}>0$, and when it is elongated, ${\delta}<0$. The orbital triplet splits into three Kramers doublets. Their eigenvalues are expressed as
\begin{eqnarray}
\frac{E_{\rm l}}{{\lambda}^{\prime}}=\frac{{\delta}}{3{\lambda}^{\prime}}+\frac{1}{2},
\label{eigen1}
\end{eqnarray}
and
\begin{eqnarray}
\frac{E_{\rm q}^{\pm}}{{\lambda}^{\prime}}=-\frac{{\delta}}{6{\lambda}^{\prime}}-\frac{1}{4}\pm\frac{1}{2}\sqrt{\left(\frac{{\delta}}{{\lambda}^{\prime}}\right)^2-\frac{{\delta}}{{\lambda}^{\prime}}+\frac{9}{4}}.
\label{eigen2}
\end{eqnarray}
These eigenvalues are shown in Fig.\,\ref{Kramers} as a function of ${\delta}/{\lambda}^{\prime}$.
 
When the temperature $T$ is much lower than ${\lambda}^{\prime}\,{\simeq}\,1000$~cm$^{-1}$,\cite{Geschwind} i.e., $T\,{<}~100$\,K, the magnetic property is determined by the lowest Kramers doublet with $E\,{=}\,E_{\rm q}^{-}$. The eigen-states of the lowest Kramers doublet are expressed as
\begin{eqnarray}
{\psi}_{\pm}=c_1|\pm 1, \mp 1/2{\rangle}+c_2|0, \pm 1/2{\rangle},
\label{state}
\end{eqnarray}
where $|m_l, m_S{\rangle}$ denotes the state with $l^z\,{=}\,m_l$ and $S^z\,{=}\,m_S$. Coefficients $c_1$ and $c_2$ are given by
\begin{eqnarray}
c_1=\frac{1}{\sqrt{2}}\sqrt{1\,{-}\,\frac{A}{\sqrt{A^2{+}1}}},\ \ 
c_2=-\frac{1}{\sqrt{2}}\sqrt{1\,{+}\,\frac{A}{\sqrt{A^2{+}1}}},
\label{coefficient}
\end{eqnarray}
with
\begin{eqnarray}
A=\frac{2({\delta}/{\lambda}^{\prime})-1}{2\sqrt{2}}.
\label{A}
\end{eqnarray}

Within the lowest Kramers doublet, we have
\begin{eqnarray}
\begin{array}{c}
{\langle}{\psi}_{\pm}|S^z|{\psi}_{\pm}{\rangle}=\mp\frac{1}{2}(c_1^2-c_2^2),\vspace{3mm}\\
{\langle}{\psi}_{+}|S^+|{\psi}_{-}{\rangle}={\langle}{\psi}_{-}|S^-|{\psi}_{+}{\rangle}=c_2^2.
\end{array}
\label{spin1}
\end{eqnarray}
Using these relations, we can replace the true spin $\bm S$ with $S\,{=}\,1/2$ by the spin-1/2 operator $\bm s$ given by
\begin{eqnarray}
S^x=c_2^2s^x,\hspace{2mm}S^y=c_2^2s^y,\hspace{2mm}S^z=-(c_1^2-c_2^2)s^z.
\label{spin2}
\end{eqnarray}
We assume that the exchange interaction between true spins ${\bm S}_i$ and ${\bm S}_j$ is described by the Heisenberg model ${\cal H}_{\rm ex}=J{\bm S}_i\cdot{\bm S}_j$. Substituting eq.~(\ref{spin2}) into ${\cal H}_{\rm ex}$, we obtain the effective model  
\begin{eqnarray}
{\cal H}_{\rm eff}=J^{\perp}\left(s_i^xs_j^x+s_i^ys_j^y\right)+J^{\parallel}s_i^zs_j^z,
\label{model}
\end{eqnarray}
with
\begin{eqnarray}
J^{\perp}=c_2^4J,\hspace{5mm}J^{\parallel}=(c_1^2-c_2^2)^2J.
\label{model2}
\end{eqnarray}
The exchange constants $J^{\perp}$ and $J^{\parallel}$ in the special cases are shown in Table \ref{table1}. When the trigonal crystalline field is absent (${\delta}\,{=}\,0$), the effective exchange interaction ${\cal H}_{\rm eff}$ becomes the Heisenberg model, while when ${\delta}/{\lambda}^{\prime}\,{=}\,1/2$, it becomes the complete XY model. For ${\delta}/{\lambda}^{\prime}\,{\rightarrow}\,{\infty}$, the orbital angular momentum is quenched, so that the magnetic moment is given by the spin only, and ${\cal H}_{\rm eff}$ again becomes the Heisenberg model. For ${\delta}/{\lambda}^{\prime}\,{<}\,0$, ${\cal H}_{\rm eff}$ is Ising-like and becomes the complete Ising model for ${\delta}/{\lambda}^{\prime}\,{\rightarrow}\,{-\infty}$.

\begin{table}[hbt]
\caption{\label{table1}Coefficients $c_1$ and $c_2$, exchange constants $J^{\perp}$ and $J^{\parallel}$ and $g$-factors $g^{\perp}$ and $g^{\parallel}$ in the special cases of ${\delta}/{\lambda}^{\prime}$.}
\begin{ruledtabular}
\begin{tabular}{ccccccc}
${\delta}/{\lambda}^{\prime}$ & $c_1$ & $c_2$ & $J^{\perp}$ & $J^{\parallel}$ & $g^{\perp}$ & $g^{\parallel}$ \\
\hline
0 & $\sqrt{2/3}$ & $-1/{\sqrt{3}}$ & $J/9$ & $J/9$ & $2(1+2k)/3$ & $2(1+2k)/3$\vspace{1mm} \\
1/2 & $1/{\sqrt{2}}$ & $-1/{\sqrt{2}}$ & $J/4$ & 0 & $1+{\sqrt{2}}k$ & $k$\vspace{1mm} \\
1 & $1/\sqrt{3}$ & $-\sqrt{2/3}$ & $4J/9$ & $J/9$ & $4(1+k)/3$ & $2(1-k)/3$\vspace{0.5mm} \\
${\infty}$ & 0 & $-1$ & $J$ & $J$ & 2 & 2 \\
$-{\infty}$ & 1 & 0 & 0 & $J$ & 0 & $2(k+1)$ \\
\end{tabular}
\end{ruledtabular}
\end{table}

\begin{figure}[tb]
\begin{center}
\includegraphics[width=0.8\linewidth]{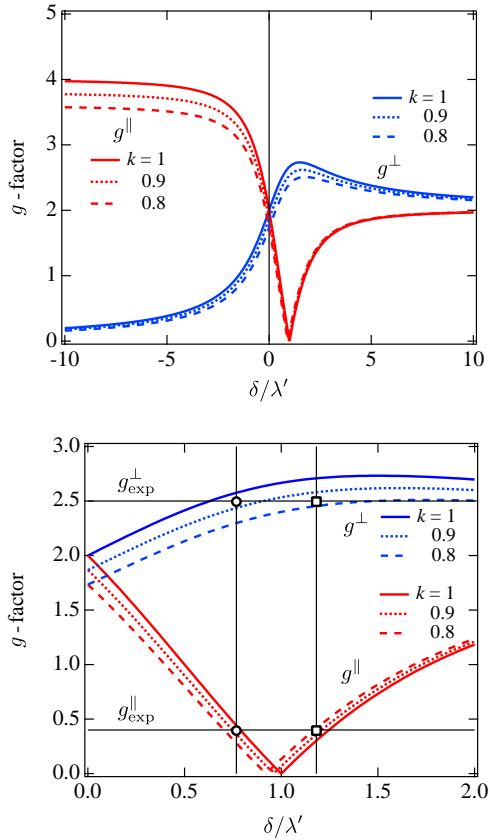}
\end{center}
\caption{(a) $g$-factors as a function of ${\delta}/{\lambda}^{\prime}$. (b) Enlargement of the $g$-factors between ${\delta}/{\lambda}^{\prime}\,{=}\,0$ and 2. The two horizontal lines are experimental $g$-factors estimated from the high-field magnetization process. Open circles and squares represent two sets of the $g$-factors suitable for $\alpha$-RuCl$_3$.} 
\label{g-factor}
\end{figure}

The lowest Kramers doublet splits into two Zeeman levels when subjected to a magnetic field. The splitting of the Zeeman levels in Ru$^{3+}$ has been discussed by many authors.\cite{Geschwind,Wu,He} The Zeeman term is written as
\begin{eqnarray}
{\cal H}_{\rm Z}=
-{\mu}_{\rm B}(-k{\bm l}+2{\bm S})\cdot{\bm H}.
\label{Zeeman}
\end{eqnarray}
When a magnetic field is applied parallel to the trigonal axis, the Zeeman energy is expressed as
\begin{eqnarray}
{\cal H}_{\rm Z}^{\parallel}=-{\mu}_{\rm B}{\langle}{\psi}_{\mp}|(-kl^z+2S^z)|{\psi}_{\mp}{\rangle}H=-g^{\parallel}{\mu}_{\rm B}s^zH,
\label{Zeeman_para}
\end{eqnarray}
with
\begin{eqnarray}
g^{\parallel}=2|\{(k+1)c_1^2-c_2^2\}|.
\label{g_para}
\end{eqnarray}
Note that the Zeeman levels of ${\psi}_{-}$ and ${\psi}_{+}$ are reversed at $(c_2/c_1)^2\,{=}\,k\,{+}\,1$.
The Zeeman energy for a magnetic field perpendicular to the trigonal axis is expressed as
\begin{eqnarray}
{\cal H}_{\rm Z}^{\perp}\hspace{-1mm}&=&\hspace{-1mm}-\frac{1}{2}{\mu}_{\rm B}{\langle}{\psi}_{\pm}|-k(l^+\,{+}\,l^-)+2(S^+\,{+}\,S^-)|{\psi}_{\mp}{\rangle}H\nonumber\\
\hspace{-1mm}&=&\hspace{-1mm}-g^{\perp}{\mu}_{\rm B}s^xH,
\label{Zeeman_perp}
\end{eqnarray}
with
\begin{eqnarray}
g^{\perp}=2(c_2^2-\sqrt{2}kc_1c_2).
\label{g_perp}
\end{eqnarray}
Figure \ref{g-factor} shows these $g$-factors as a function of ${\delta}/{\lambda}^{\prime}$. The $g$-factors in the special cases are shown in Table \ref{table1}.

The $g$-factors estimated from the high-field magnetization process are $g_{ab}\,{=}\,g_{\rm exp}^{\perp}\,{\simeq}\,2.5$ and $g_{c}\,{=}\,g_{\rm exp}^{\parallel}\,{\simeq}\,0.4$. This indicates that ${\delta}/{\lambda}^{\prime}\,{\sim}\,1$ in RuCl$_3$. Figure \ref{g-factor}(b) shows the behavior of the $g$-factors in the range of $0\leq{\delta}/{\lambda}^{\prime}\leq2$, where $g^{\parallel}$ changes rapidly with varying ${\delta}/{\lambda}^{\prime}$. There are two sets of parameters, $({\delta}/{\lambda}^{\prime}, k)\,{=}\,(0.77,\ 0.95)$ and $(1.18,\ 0.83)$, that satisfy the $g$-factors observed for $\alpha$-RuCl$_3$. In the present experiment, it is difficult to evaluate which set of parameters is realized for RuCl$_3$. The exchange anisotropy is given as $J^{\parallel}/J^{\perp}\,{=}\,0.099$ and 0.37 for ${\delta}/{\lambda}^{\prime}\,{=}\,0.77$ and 1.18, respectively. Hence, the exchange interaction between effective spins is strongly XY-like in $\alpha$-RuCl$_3$. From the above discussion, we can conclude that the strongly anisotropic magnetic properties observed in $\alpha$-RuCl$_3$ arise from the trigonal crystalline field, which is close to the spin-orbit coupling, ${\delta}/{\lambda}^{\prime}\,{\simeq}\,1$. When uniaxial pressure is applied parallel to the trigonal axis, that is, normal to the $ab$ plane, the coefficient $\delta$ of the trigonal crystalline field will increase because the magnitude of the trigonal compression of the RuCl$_6$ octahedron is increased by the uniaxial pressure. If ${\delta}/{\lambda}^{\prime}\,{=}\,0.77$ at ambient pressure, $g_c$ will decrease under the uniaxial pressure, while if ${\delta}/{\lambda}^{\prime}\,{=}\,1.18$, $g_c$ will increase. Thus, magnetization measurements under the uniaxial pressure will be useful in determining the parameter ${\delta}/{\lambda}^{\prime}$.

\subsection{Successive magnetic phase transitions}

As shown in the phase diagram of Fig.\,\ref{phase}, $\alpha$-RuCl$_3$ undergoes four magnetic phase transitions at zero magnetic field. If the frustration is absent, then a single magnetic phase transition will occur because the honeycomb lattice is bipartite. Hence, it is considered that the successive phase transitions arise from the frustration effect owing to the competing interactions.
 
Recently, Sears {\it et al.}\cite{Sears} investigated the spin structure below $T_{\rm N4}\,{=}\,7.6$ K at zero magnetic field by neutron diffraction. They reported that the so-called zigzag-type order is realized below $T_{\rm N4}$. They discussed the ground state from the viewpoint of the competition between the Heisenberg term $J{\bm S}_i\,{\cdot}\,{\bm S}_j$ and the Kitaev term $-KS_i^{\gamma}S_j^{\gamma}$, where $\gamma$ corresponds to the direction of the bond connecting ${\bm S}_i$ and ${\bm S}_j$. A possible origin of the Kitaev term is discussed in Appendix. However, it appears difficult to derive the successive phase transitions within the nearest neighbor interactions. 

In MnBr$_2$, MnI$_2$ and NiBr$_2$, which have the similar layered crystal structure to $\alpha$-RuCl$_3$, two magnetic phase transitions have been observed at zero magnetic field.\cite{Iio, Sato,Day,Adam} Their successive phase transitions have been explained theoretically in terms of the competition among the exchange interactions up to the third neighbor and the interlayer exchange interaction.\cite{Yoshiyama1,Yoshiyama2} 

When MX$_6$ octahedra centered by magnetic ions M form a honeycomb or triangular lattice by sharing their edges, the nearest-neighbor exchange bond angle M$-$X$-$M is close to 90$^{\circ}$. In $\alpha$-RuCl$_3$, the nearest-neighbor exchange bond angle Ru$-$Cl$-$Ru is approximately 96$^{\circ}$. When the bond angle is close to 90$^{\circ}$, the exchange tends to be ferromagnetic or weak, even if it is antiferromagnetic.\cite{Kanamori} Actually, in closely related CrCl$_3$, a ferromagnetic ordering is realized in the honeycomb lattice.\cite{Mizette,Kuhlow} The second- and third-neighbor exchange interactions $J_2$ and $J_3$ in $\alpha$-RuCl$_3$ are considered to be the same order of magnitude as the nearest-neighbor exchange interaction $J_1$, as observed in MnBr$_2$, MnI$_2$ and NiBr$_2$. 

The classical ground state for the $J_1\,{-}\,J_2\,{-}\,J_3$ Heisenberg and XY model was theoretically investigated by Rastelli {\it et al.}\cite{Rastelli} and Fouet {\it et al.}\cite{Fouet} They showed that zigzag ordering emerges when $J_2/J_1\,{>}\,1/2$ and $J_3/J_1\,{>}\,1/2$ for antiferromagnetic $J_1\,({>}\,0)$ and when $J_2/J_1\,{<}\,1/2$ and $J_3/J_1\,{<}\,0$ for ferromagnetic $J_1\,({<}\,0)$. Thus, zigzag ordering is possible in a realistic parameter range.
Therefore, we infer that the successive phase transitions and zigzag magnetic ordering observed in $\alpha$-RuCl$_3$ are attributed to the competition among the exchange interactions up to the third neighbor and the interlayer exchange interaction. However, a theoretical description of the successive phase transitions is an open problem.

\section{Conclusion}

We have presented the results of magnetization and specific heat measurements on the honeycomb-lattice magnet $\alpha$-RuCl$_3$. This compound undergoes a first-order structural phase transition at $T_{\rm t}\,{=}\,154\,{\pm}\,13$ K. The structural phase transition is expected to be a transition from the monoclinic room-temperature structure ($C2/m$) to a trigonal structure by analogy with that observed in closely related CrCl$_3$.\cite{Morosin} 

The magnetic susceptibility and magnetization are strongly anisotropic, i.e., these quantities for $H\,{\parallel}\,ab$ are much larger than those for $H\,{\perp}\,ab$. This is ascribed to the strongly anisotropic $g$-factor characteristic of the low-spin state of Ru$^{3+}$ with the $4d^5$ electronic state. We discussed the effective exchange interaction and $g$-factor taking the spin-orbit coupling and trigonal crystalline field into consideration. We demonstrated that the strongly anisotropic magnetic properties observed in $\alpha$-RuCl$_3$ occur when the magnitudes of the spin-orbit coupling and trigonal crystalline field are close to each other, i.e., ${\delta}/{\lambda}^{\prime}\,{\simeq}\,1$, and that the effective exchange interaction is strongly XY-like in $\alpha$-RuCl$_3$. 

It was found from the specific heat and magnetization measurements that $\alpha$-RuCl$_3$ undergoes four magnetic phase transitions at zero magnetic field and five field-induced transitions at $T\,{=}\,0$. We presented the magnetic field vs temperature phase diagram for $H\,{\parallel}\,ab$ in Fig.\,\ref{phase}. We suggest that these successive phase transitions are attributed to the competition among the nearest-neighbor, second- and third-neighbor exchange interactions.

\begin{acknowledgments}
We express our profound gratitude to H. Uekusa for useful discussion on the crystal structure. This work was supported by Grants-in-Aid for Scientific Research from Japan Society for the Promotion of Science (A) Nos. 17204028 and 26247058.
\end{acknowledgments}

\appendix

\section{Possible origin of the Kitaev term}
\renewcommand{\thefigure}{A\arabic{figure}}

\begin{figure}[t]
\begin{center}
\includegraphics[width=0.70\linewidth]{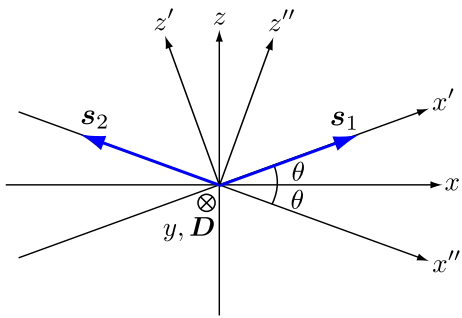}
\end{center}
\caption{Classical configuration of spins ${\bm s}_i$ and ${\bm s}_j$ and coordinate systems. The $y$ direction is chosen to be parallel to the bond vector ${\bm r}_{12}$ and ${\bm D}$.}
\label{Kitaev}
\end{figure}

Here, we give a possible origin of the Kitaev term. As discussed above, the dominant exchange interaction between magnetic moments is the spin-1/2 XXZ model expressed by eq.\,(\ref{model}), which is isotropic in the $ab$ plane. We consider the Dzyaloshinskii-Moriya (DM) interaction ${\bm D}\,{\cdot}\,[{\bm s}_1{\times}{\bm s}_2]$ with the ${\bm D}$ vector parallel to the bond vector ${\bm r}_{12}\,{=}\,{\bm r}_2\,{-}\,{\bm r}_1$. This condition is allowed when the space group is $P3_112$. In this case, there is a twofold axis passing two neighboring spins, which leads to ${\bm D}\,{\parallel}\,{\bm r}_{12}$.\cite{Moriya} The interaction between the nearest-neighbor spins ${\bm s}_1$ and ${\bm s}_2$ is expressed as
\begin{eqnarray}
{\cal H}_{12}=J^{\perp}\hspace{-0.5mm}\left(s_1^xs_2^x\,{+}\,s_1^ys_2^y\right)+J^{\parallel}s_1^zs_2^z+{\bm D}\,{\cdot}\left[{\bm s}_1{\times}{\bm s}_2\right],
\label{model_A1}
\end{eqnarray}
where the $z$ axis is chosen to be normal to the honeycomb lattice.
When the XXZ-type exchange interaction is antiferromagnetic, the stable classical spin configuration is a canted antiferromagnetic state as illustrated in Fig.~\ref{Kitaev}. The canting angle $\theta$ is given by 
\begin{eqnarray}
\tan{2\theta}=\frac{2D}{J^{\perp}+J^{\parallel}}.
\end{eqnarray}

Here, we define the local coordinates $x^{\prime}yz^{\prime}-$O and $x^{\prime\prime}yz^{\prime\prime}-$O as shown in Fig.~\ref{Kitaev}, i.e., the $x^{\prime}$ and $x^{\prime\prime}$ axes are taken to be parallel and antiparallel to the spins ${\bm s}_1$ and ${\bm s}_2$, respectively. The $y$ axis is parallel to the bond vector and $\bm D$. The spin operators $s_{1,2}^x$ and $s_{1,2}^z$ in the original coordinate system are expressed as 
\begin{eqnarray}
s_1^x=s_1^{x^{\prime}}\hspace{-0.5mm}\cos{\theta}-s_1^{z^{\prime}}\hspace{-0.5mm}\sin{\theta},\hspace{2mm}
s_1^z=s_1^{x^{\prime}}\hspace{-0.5mm}\sin{\theta}+s_1^{z^{\prime}}\hspace{-0.5mm}\cos{\theta},\nonumber\\
s_2^x=s_2^{x^{\prime\prime}}\hspace{-1mm}\cos{\theta}+s_2^{z^{\prime\prime}}\hspace{-1mm}\sin{\theta},\hspace{2mm}
s_2^z=-s_2^{x^{{\prime}{\prime}}}\hspace{-1mm}\sin{\theta}+s_2^{z^{{\prime}{\prime}}}\hspace{-1mm}\cos{\theta}.\nonumber\\
\ 
\label{transformation}
\end{eqnarray}
Substituting eq.\,(\ref{transformation}) into eq.\,(\ref{model_A1}), we obtain
\begin{eqnarray}
{\cal H}_{12}=&&\hspace{-6mm}(J^{\perp}+K)\hspace{-0.8mm}\left(s_1^{x^{\prime}}\hspace{-0.6mm}s_2^{x^{{\prime}{\prime}}}\,{+}\,s_1^y\hspace{0.3mm}s_2^y\right)+(J^{\parallel}+K)s_1^{z^{\prime}}\hspace{-0.6mm}s_2^{z^{{\prime}{\prime}}}\nonumber\\
&-&\hspace{-1mm}Ks_1^y\hspace{0.3mm}s_2^y,
\label{model_A2}
\end{eqnarray}
with
\begin{eqnarray}
K=2D\sin{\theta}\cos{\theta}-(J^{\perp}+J^{\parallel})\sin^2{\theta}.
\end{eqnarray}
Because the $y$ axis is parallel to the bond vector ${\bm r}_{12}$, which has three directions in the crystal depending on the configuration of ${\bm s}_1$ and ${\bm s}_2$, eq.\,(\ref{model_A2}) is equivalent to the spin-1/2 Kitaev-XXZ model. As described above, the DM interaction can be the origin of the Kitaev term.

\end{document}